\documentclass[conference]{IEEEtran}
\IEEEoverridecommandlockouts
\usepackage{cite}
\usepackage{amsmath,amssymb,amsfonts}
\usepackage{algorithmic}
\usepackage{graphicx}
\usepackage{textcomp}
\usepackage{xcolor}
\def\BibTeX{{\rm B\kern-.05em{\sc i\kern-.025em b}\kern-.08em
    T\kern-.1667em\lower.7ex\hbox{E}\kern-.125emX}}

\usepackage{listings}
\usepackage{tikz}
\usepackage{pgfplots}
\usepackage{amssymb}
\usepackage{float}
\usepackage{graphicx}

\usepackage[utf8]{inputenc}
\usepackage{adjustbox}
\DeclareUnicodeCharacter{2212}{−}
\usepgfplotslibrary{groupplots,dateplot}
\usetikzlibrary{patterns,shapes.arrows}
\pgfplotsset{compat=newest}
\pgfplotsset{compat=1.18}
\usepackage{xintexpr}
\usepackage{multirow}
\usepackage{longtable}
\usepackage{url}
\usepackage{booktabs}
\usepackage{amsmath}
\usepackage[ruled, vlined, linesnumbered]{algorithm2e}
\usepackage{subcaption}
\usepackage{makecell}

\newcommand{\vc}{{\mathbf{c}}}
\newcommand{\vd}{{\mathbf{d}}}
\newcommand{\ve}{{\mathbf{e}}}
\newcommand{\vf}{{\mathbf{f}}}

\newcommand{\vp}{{\mathbf{p}}}
\newcommand{\vq}{{\mathbf{q}}}
\newcommand{\vr}{{\mathbf{r}}}

\newcommand{\vu}{{\mathbf{u}}}

\newcommand{\vw}{{\mathbf{w}}}

\newcommand{\vy}{{\mathbf{y}}}

\newcommand{\valpha}{{\boldsymbol{\alpha}}}
\newcommand{\vlambda}{{\boldsymbol{\lambda}}}

\newcommand{\mB}{{\mathbf{B}}}

\newcommand{\mF}{{\mathbf{F}}}
\newcommand{\mG}{{\mathbf{G}}}

\newcommand{\mI}{{\mathbf{I}}}

\newcommand{\mK}{{\mathbf{K}}}
\newcommand{\mL}{{\mathbf{L}}}
\newcommand{\mM}{{\mathbf{M}}}

\newcommand{\mO}{{\mathbf{O}}}
\newcommand{\mP}{{\mathbf{P}}}

\newcommand{\mR}{{\mathbf{R}}}

\newcommand{\mU}{{\mathbf{U}}}

\DeclareMathOperator{\Kernelof}{Ker}
    
\begin{document}

\title{Assembly of FETI dual operator using CUDA
}

\author{\IEEEauthorblockN{Jakub Homola, Radim Vav\v{r}\'{i}k, Ond\v{r}ej Meca, Tom\'{a}\v{s} Brzobohat\'{y}, and Lubom\'{i}r \v{R}\'{i}ha}
\IEEEauthorblockA{\textit{IT4Innovations}, 
\textit{VSB – Technical University of Ostrava}
Ostrava-Poruba, Czech Republic \\
jakub.homola@vsb.cz, radim.vavrik@vsb.cz, ondrej.meca@vsb.cz, tomas.brzobohaty@vsb.cz, and lubomir.riha@vsb.cz}
}


\maketitle
\thispagestyle{plain}
\pagestyle{plain}

\begin{abstract}

FETI is a numerical method used to solve engineering problems.
It builds on the ideas of domain decomposition, which makes it highly scalable and capable of efficiently utilizing whole supercomputers.
One of the most time-consuming parts of the FETI solver is the application of the dual operator $\mF$ in every iteration of the solver.

It is traditionally performed on the CPU using an implicit approach of applying the individual sparse matrices that form $\mF$ right-to-left.
Another approach is to apply the dual operator explicitly, which primarily involves a simple dense matrix-vector multiplication and can be efficiently performed on the GPU.
However, this requires additional preprocessing on the CPU where the dense matrix is assembled, which makes the explicit approach beneficial only after hundreds of iterations are performed.

In this paper, we use the GPU to accelerate the assembly process as well. This significantly shortens the preprocessing time, thus decreasing the number of solver iterations needed to make the explicit approach beneficial.

With a proper configuration, we only need a few tens of iterations to achieve speedup relative to the implicit CPU approach.
Compared to the CPU-only explicit approach, we achieved up to 10$\times$ speedup for the preprocessing and 25$\times$ for the application.

\end{abstract}


\section{Introduction}

FETI methods (finite element tearing and interconnecting) are highly scalable numerical methods used to solve large-scale engineering problems. It builds on the principles of the finite element method (FEM) and domain decomposition. The spatial domain is decomposed into several smaller subdomains, which are glued together using Lagrange multipliers to enforce equality on the interfaces between them. The resulting system is usually solved using iterative methods, where in each iteration, subdomains can be solved concurrently. The convergence properties are, in a sense, optimal~\cite{FETI_OPT,FETI_CONVERGENCE}.

With the increasing power of modern high-performance clusters, many systems are based on GPUs since they offer much higher FLOP/s performance, memory bandwidth, and energy efficiency compared to CPUs.
Hence, it is tempting to utilize GPUs for FETI algorithms as well.

It is natural to try to accelerate the most time-consuming part of an algorithm. In the FETI solver, a significant portion of the runtime is taken up by the application of the dual operator $\mF = \mB \mK^{+} \mB^\top$ in every iteration of the solver.

The application of $\mF$ is traditionally performed on the CPU using the implicit approach of applying the individual sparse matrices right-to-left.
The other option is to use the explicit approach. There, the dense matrices that form $\mF$ need to be assembled first. The application is then just a simple dense matrix-vector multiplication. This approach significantly speeds up iterations of the FETI solver at the cost of expensive preprocessing, where the assembly takes place.

Because the dense matrix-vector multiplication kernel is very well suited for GPUs, previous attempts at acceleration used the explicit approach. They assembled the matrices that form $\mF$ on the CPU and copied them to the accelerator only for the application~\cite{BDDS_ACC,FETI_PHI,ESPRESO-SC}.

Some of them used a more clever algorithm for assembly -- an augmented incomplete factorization method from the PARDISO library~\cite{pardiso}. It takes advantage of the high sparsity of $\mB$ to speed up the assembly.
However, this approach is still beneficial only for problems requiring thousands of iterations of the FETI solver (e.g., ill-conditioned problems).

In this work, we also use the explicit approach. However, contrary to all of the mentioned acceleration attempts, we use the GPU to assemble the local dual operator as well. As shown, this approach is faster than the augmented incomplete factorization on the CPU for most non-trivial 3D problems.

There are multiple ways in which the explicit local dual operator can be assembled. We can choose between cuBLAS and cuSPARSE kernels; we can call each of them with different parameters (e.g., use the transpose flag or transpose the matrix manually) or even use a slightly different algorithm. The best parameter setting depends on the type of solved problem (2D or 3D), the size of the subdomain, and, surprisingly, the version of CUDA libraries. In this paper, we provide a summary of the optimal parameter settings, which is based on an exhaustive search of the parameter space.

With the appropriate settings, we achieved up to $10\times$ speedup of the explicit assembly
and $25\times$ speedup of the application,
both relative to the Intel MKL PARDISO CPU-only explicit approach (Karolina GPU node -- 8x Nvidia A100 GPUs, 2x 64-core AMD EPYC 7763 CPUs).
With this, we reduce the number of iterations required for the GPU approach to become beneficial (the amortization point) from hundreds to tens.
For problems requiring hundreds of iterations, we observed a speed-up of around 10 (relative to the traditional CPU implicit approach) in the dual operator-related parts of the FETI solver.

The paper is structured as follows.
Section~\ref{sec: theory} presents the basic theory of FETI algorithms focusing on the FETI dual operator.
Section~\ref{sec:implementation} describes key aspects of the implementation and efficient parallelization of FETI methods needed to fully understand the acceleration.
Section~\ref{sec:acceleration} describes the acceleration of explicit evaluation of $\mF$.
Section~\ref{sec:performance} compares different approaches and summarizes achieved results.

\section{Finite Element Tearing and Interconnecting}
\label{sec: theory}


Let us first quickly introduce the FETI method~\cite{FETI}. It is a modification of the finite element method (FEM)~\cite{fem}, which is a numerical method for solving differential equations. FETI uses domain decomposition techniques to allow solving of FEM on large-scale parallel systems, which is its main advantage over FEM. 
For readers unfamiliar with FETI, we provide a short description.

We start with the FEM system
\begin{equation}
    \label{eq:femsystem}
    \tilde{\mK} \tilde{\vu} = \tilde{\vf},
\end{equation}
where $\tilde{\mK}$ is the stiffness matrix, $\tilde{\vf}$ is the load vector, and $\tilde{\vu}$ is the solution.

By rearranging the order of DOFs and adding several carefully crafted equations and variables, we are able to transform the FEM system into an equivalent system with a block structure,
\begin{equation}
    \label{eq:fetisystemblocked}
    \begin{bmatrix}
        \mK_1 & & & & \mB_1^\top \\
        & \mK_2 & & & \mB_2^\top \\
        & & \ddots & & \vdots \\
        & & & \mK_N & \mB_N^\top \\
        \mB_1 & \mB_2 & \dots & \mB_N & \mO
    \end{bmatrix}
    \begin{bmatrix}
        \vu_1 \\
        \vu_2 \\
        \vdots \\
        \vu_N \\
        \vlambda
    \end{bmatrix}
    =
    \begin{bmatrix}
        \vf_1 \\
        \vf_2 \\
        \vdots \\
        \vf_N \\
        \vc
    \end{bmatrix}
\end{equation}
or, shortly,
\begin{equation}
    \label{eq:fetisystem}
    \begin{bmatrix}
        \mK & \mB^\top \\
        \mB & \mO
    \end{bmatrix}
    \begin{bmatrix}
        \vu \\
        \vlambda
    \end{bmatrix}
    =
    \begin{bmatrix}
        \vf \\
        \vc
    \end{bmatrix}
\end{equation}
where each block represents one of the $N$ subdomains into which we decomposed the spatial domain.
$\mK$ is the block diagonal stiffness matrix comprising of the subdomain stiffness matrices $\mK_i$, $\vlambda$ is the Lagrange multiplier vector (we call all vectors with size equal to $\vlambda$ \textit{dual vectors}), and the matrix $\mB$ represents the subdomain gluing.
We use the Total FETI variant~\cite{totalfeti}, where we isolate the Dirichlet boundary conditions and append them to the gluing matrix $\mB$ and the right-hand side $\vc$. This makes all the subdomain stiffness matrices singular.
Such a rearrangement and expansion of the system makes the matrix larger, but it makes the parallelization of the solution more straightforward.

We denote $\mR_i$ to be the matrix containing the basis vectors of $\Kernelof \mK_i$ in its columns, and we create $\mR$ by placing $\mR_i$ on the main block diagonal.
From the solvability of the first equation in \eqref{eq:fetisystem}, we eventually get
\begin{equation}
    \label{eq:solvability}
    - \mR^\top \mB^\top \vlambda = - \mR^\top \vf.
\end{equation}
Given the Lagrange multipliers $\vlambda$, we can express $\vu$ using
\begin{equation}
    \label{eq:solutionUeval}
    \vu = \mK^+ (\vf - \mB^\top \vlambda) + \mR \valpha,
\end{equation}
where $\mK^+$ is a generalized inverse of $\mK$ satisfying $\mK \mK^+ \mK = \mK$.
Using \eqref{eq:solutionUeval} in the second equation in \eqref{eq:fetisystem}, we eventually get
\begin{equation}
    \label{eq:alphaconstraint}
    \mB \mK^+ \mB^\top \vlambda - \mB \mR \valpha = \mB \mK^+ \vf - \vc.
\end{equation}
We define $\mF = \mB \mK^+ \mB^\top$ (the \textit{dual operator}), $\mG = \mB \mR$, $\vd = \mB \mK^+ \vf - \vc$, $\ve = \mR^\top \vf$ and combine \eqref{eq:solvability} with \eqref{eq:alphaconstraint}, and get
\begin{equation}
    \label{eq:tfetidualproblem}
    \begin{bmatrix}
        \mF & -\mG \\
        -\mG^\top & \mO
    \end{bmatrix}
    \begin{bmatrix}
        \vlambda \\
        \valpha
    \end{bmatrix}
    =
    \begin{bmatrix}
        \vd \\
        -\ve
    \end{bmatrix}.
\end{equation}

The system \eqref{eq:tfetidualproblem} is solved using the preconditioned conjugate projected gradient method (PCPG), using the projector
\begin{equation}
    \mP = \mI - \mG (\mG^\top \mG)^{-1} \mG^\top.
\end{equation}
and a preconditioner $\mM$. We use Algorithm~\ref{alg:fetipcpg} to find $\vlambda$, $\valpha$ can be then computed using
\begin{equation}
\label{eq:alphacompute}
    \valpha = - (\mG^\top \mG)^{-1} \mG^\top (\vd - \mF \vlambda).
\end{equation}

\begin{algorithm}
\caption{The preconditioned conjugate projected gradient method}
\label{alg:fetipcpg}
\DontPrintSemicolon
$\vlambda_0 \gets \vlambda_I$ \;
$\vr_0 \gets \vd - \mF \vlambda_I$ \;
$\vw_0 \gets \mP \vr_0$ \;
$\vy_0 \gets \mP \mM \vw_0$ \;
$\vp_0 \gets \vy_0$ \;
\For{$k = 0, 1, 2, \dots$ until convergence}{
    $\vq_k \gets \mF \vp_k$ \;
    $\delta_k \gets (\vw_k^\top \vy_k) / (\vp_k^\top \vq_k)$ \;
    $\vlambda_{k+1} \gets \vlambda_k + \delta_k \vp_k$ \;
    $\vr_{k+1} \gets \vr_k - \delta_k \vq_k$ \;
    $\vw_{k+1} \gets \mP \vr_{k+1}$ \;
    $\vy_{k+1} \gets \mP \mM \vw_{k+1}$ \;
    $\beta_k \gets (\vw_{k+1}^\top \vy_{k+1}) / (\vw_k^\top \vy_k)$ \;
    $\vp_{k+1} \gets \vy_{k+1} + \beta_k \vp_k$ \;
}
\end{algorithm}

\subsection{Dual operator $\mF$}
\label{sec:dualop}

The matrix $\mF$, defined previously as
\begin{equation}
    \mF = \mB \mK^+ \mB^\top,
\end{equation}
is sometimes called the dual operator, dual Schur complement matrix, or the FETI operator. 
It is square, and each row and column corresponds to a single Lagrange multiplier (gluing connection between subdomains).
Due to the block structure of the matrices, we are able to define a \textit{local dual operator} for each subdomain as
\begin{equation}
    \label{eq:localdualoperator}
    \tilde{\mF}_i = \tilde{\mB}_i \mK_i^+ \tilde{\mB}_i^\top.
\end{equation}
It is a (possibly non-contiguous) submatrix of $\mF$ defined by only the Lagrange multipliers connected to the $i$-th subdomain. The individual $\tilde{\mF}_i$ can be combined additively to form $\mF$.

The application of the dual operator $\mF$, therefore, primarily consists of the (possibly concurrent) application of the local dual operator $\tilde{\mF}_i$ for each subdomain.

\section{Implementation of FETI solver}
\label{sec:implementation}

This section describes the fundamental aspects of the FETI solver w.r.t. the dual operator and its main constraints for porting it to the GPU accelerators. 
The fundamental principle of FETI algorithms is to divide the workload and distribute it among available hardware resources.
With current architectures, a good distribution usually takes into account both shared and distributed memory.

\begin{figure}
 \centering
 \includegraphics[width=\columnwidth]{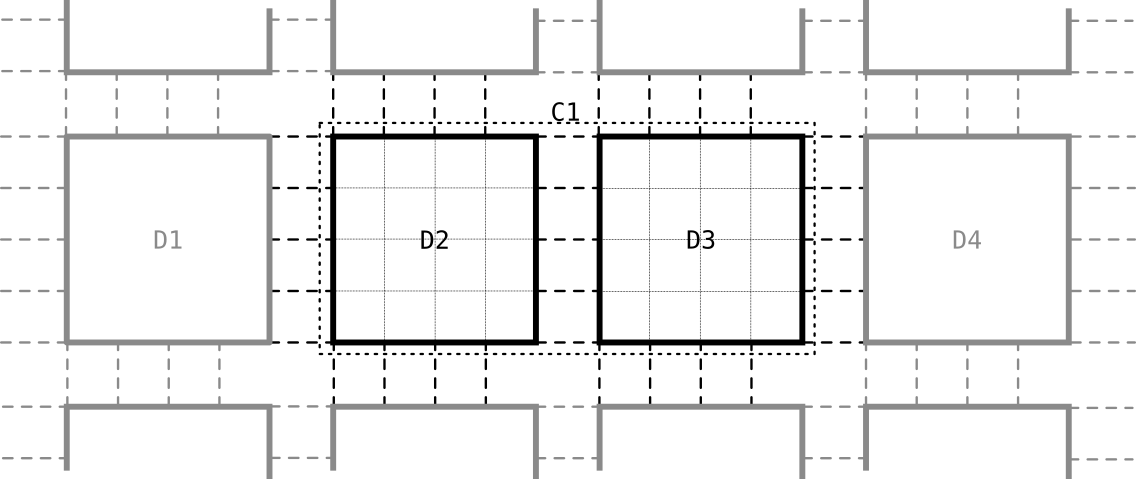}
 \caption{Example of domain decomposition and clusterization of subdomains to clusters. Clusterization allows a hybrid parallelization for both shared and distributed memory systems.}
 \label{fig:domains}
\end{figure}

One of the possible decompositions of a domain into subdomains is shown in Fig.~\ref{fig:domains}.
The figure shows several subdomains and their gluings to neighboring subdomains (the Lagrange multipliers), which are drawn with dashed lines. 
Several subdomains are grouped into a \textit{cluster} of subdomains (in the figure, subdomains D2 and D3 form the cluster C1, as highlighted by the dotted line).
Each process (distributed memory) then handles a single cluster.
In each cluster, subdomains are handled by threads (sharing the memory).
Typically, the number of subdomains per cluster is an integer multiple of the number of threads in order to utilize the available cores equally and achieve optimal performance of the FETI solver~\cite{espreso-pasc}.
Traditionally, each subdomain stores only the Lagrange multipliers that are connected to it.
With the introduction of clusters, Lagrange multipliers are stored per cluster.

The generalized inverse $\mK_i^+$ in \eqref{eq:localdualoperator} can be expressed as $\mK_i^+ = \mK_{i,reg}^{-1} = (\mL_i \mU_i)^{-1} = \mU_i^{-1} \mL_i^{-1}$, where $\mK_{i,reg}$ is the regularized subdomain stiffness matrix that can be obtained, e.g., through analytic regularization~\cite{brzyngeninverse}.
The local dual operator can, therefore, be written as
\begin{equation}
    \label{eq:localdualoperatorwithU}
    \tilde{\mF}_i = \tilde{\mB}_i \mU_i^{-1} \mL_i^{-1} \tilde{\mB}_i^\top.
\end{equation}

It can be applied implicitly, where individual matrices are applied from right to left,
\begin{equation}
\label{eq:dualop_apply_impl}
    \tilde{\vq}_i
    =
    \tilde{\mF}_i \tilde{\vp}_i
    =
    \tilde{\mB}_i (\mU_i^{-1} (\mL_i^{-1} (\tilde{\mB}_i^\top \tilde{\vp}_i))),
\end{equation}
or explicitly, where all matrices are multiplied at first and then applied at once,
\begin{equation}
\label{eq:dualop_apply_expl}
    \tilde{\vq}_i
    =
    \tilde{\mF}_i \tilde{\vp}_i
    =
    (\tilde{\mB}_i \mU_i^{-1} \mL_i^{-1} \tilde{\mB}_i^\top) \tilde{\vp}_i.
\end{equation}

The implicit application consists of performing a sparse matrix-vector multiplication (SPMV), 
followed by two triangular solves (TRSV) and another sparse matrix-vector multiplication.
The explicit application is a fast dense matrix-vector multiplication (GEMV), but the dense matrix $\tilde{\mF}_i$ needs to be assembled first, which can be done, e.g., with two triangular solves with a right-hand-side matrix (TRSM) and a single sparse-dense matrix multiplication (SPMM).
Comparing both approaches, \textbf{applying $\tilde{\mF}_i$ explicitly is typically faster, but the explicit assembly is a highly expensive operation, which makes the explicit approach beneficial only after a sufficient number of iterations is performed}. The amortization point (the number of iterations after which the explicit approach is faster) depends on the time spent in the explicit assembly and the time saved by the faster application.

The explicit assembly can also be done using a more sophisticated algorithm. The matrix $\tilde{\mF}_i$ can be expressed as the negative of a Schur complement of the matrix
\begin{equation}
\label{eq:matrix_for_sc}
    \begin{bmatrix}
        \mK_{i,reg} & \tilde{\mB}_i^\top \\
        \tilde{\mB}_i & \mO
    \end{bmatrix}.
\end{equation}
With this, we can use an augmented incomplete factorization designed and optimized for evaluation of the Schur complement of sparse matrices~\cite{pardiso}.

In both implicit and explicit approaches, a factorization $\mK_{i,reg} = \mL_i \mU_i$ must be performed. There are many libraries available for this task~\cite{ssolvers}.
To handle a sparsity pattern efficiently, they typically divide the factorization of the matrix into two stages~\cite{Csparse} -- symbolic and numerical. During the symbolic factorization, they search for a permutation of the matrix such that the resulting factors $\mL_i$ and $\mU_i$ have the number of non-zeros as low as possible (minimizing fill-in), and they also create the non-zero pattern of the factor. In the numerical factorization, the factor is computed and filled with values.

\begin{algorithm}
\caption{Skeleton of a multi-step simulation w.r.t. dual operators $\tilde{\mF}_i$.}
\label{alg:solverdualopview}
\DontPrintSemicolon
preparation (symbolic fact., memory buffers,...) \;
\For{every step}{
    FETI preprocessing (num. fact., assembling $\tilde{\mF}_i)$ \;
    \For{each PCPG iteration} {
        scatter Lagrange multipliers \;
        Apply $\tilde{\vq}_i = \tilde{\mF}_i \tilde{\vp}_i$ for each subdomain \;
        gather Lagrange multipliers \;
    }
}
\end{algorithm}

Algorithm~\ref{alg:solverdualopview} sketches a multi-step simulation from the dual operator point of view. It shows when the above-described routines are called.

In our use case, the structure of the finite element mesh does not change between time steps, so the non-zero pattern of the matrices $\mK_i$ also stays constant. It is, therefore, enough to call symbolic factorization only once at the beginning, in the preparation phase.

However, the non-zero values of $\mK_i$ do change between time steps, the FETI preprocessing must, therefore, be repeated. In the case of the explicit approach, this also includes reassembling the local dual operator matrices $\tilde{\mF}_i$.

After FETI preprocessing, the dual operator is ready for application in the PCPG iterations. There, the cluster-wide dual vector first needs to be \textit{scattered} into the individual subdomain-wide dual vectors $\tilde{\vp}_i$. After that, the local dual operators $\tilde{\mF}_i$ are applied, and the subdomain results $\tilde{\vq}_i$ are \textit{gathered} back into the cluster-wide dual vector. Finally, the processes synchronize their dual vectors to incorporate the changes from the neighboring clusters.

\section{The local dual operator on GPU}
\label{sec:acceleration}

This section describes the explicit assembly of the local dual operator using GPU accelerators.
We restrict ourselves to an approach based on routines provided by accelerated versions of standard high-performance mathematical libraries, such as cuBLAS or cuSPARSE~\cite{cudalibraries}, and avoid any higher-level GPU libraries.
Hence, our approach can be easily ported to a different architecture.

The local dual operator $\tilde{\mF}_i$ is assembled from the matrix $\tilde{\mB}_{i}$ and factors of $\mK_i$ using triangular solve and matrix multiplication routines.
Despite the apparent simplicity, an inappropriate algorithm can significantly degrade computational time or have unnecessarily high memory demands. The next sections describe both aspects in detail.

We use a one-to-one mapping between GPUs and clusters. This simplifies the implementation, as each process controls only a single GPU that is not shared with other processes. We will now focus on a single process, all other processes behave equally.

\subsection{Management of GPU memory}

In general, the maximum size of a problem solved by FETI on a single CPU (or GPU) is limited by the available memory.
To solve as large problems as possible, one must approach memory management wisely, especially in the case of a GPU, which typically has lower memory capacity than a CPU.
GPU memory allocations should be avoided in the hot loop because they typically cause synchronization delay and add memory management overhead.

We mentally split the GPU memory into two parts -- persistent and temporary. All persistent memory is allocated in the preparation phase and deallocated at the end of the program.
It holds mainly the structures that are not too large to fill up the memory and their size or content does not change between steps.
This includes the factors
$\mL_i$ and $\mU_i$, matrices $\tilde{\mB}_i$ and $\tilde{\mF}_i$ and the dual vectors. Some GPU libraries also require workspace buffers that need to be allocated as long as they are used, i.e., for the lifetime of the solver instance.

The rest of the memory is allocated for use in our allocator, which manages the temporary memory. In our usage, the phrase \textit{temporary memory} refers to memory buffers that are needed only for the duration of a specific kernel. These buffers are typically large, and if they were allocated persistently, the GPU memory capacity would not suffice. The temporary memory allocator is able to reuse memory without a need to call the GPU library's memory allocation routines. If there is enough remaining memory in the allocator's memory pool, memory is assigned and returned right away. Otherwise, the allocating thread is blocked until enough memory becomes available -- until other threads deallocate a sufficient amount of memory.

\subsection{Execution of the explicit dual operator stages}

As we showed in Algorithm~\ref{alg:solverdualopview}, there are three points in the program where the dual operator structures are used -- preparation, preprocessing, and application. We already briefly described them. Now, we will go into more detail.

\paragraph{Preparation} The main part of preparation is a parallel loop iterating over all subdomains in the cluster. For each subdomain, we perform the symbolic factorization, allocate the persistent structures in GPU memory, copy constant data (e.g. $\tilde{\mB}_i$ and the non-zero pattern of the factors), and call the analysis phase of cuSPARSE kernels. After the loop, we allocate the remaining memory for the temporary memory allocator.

\paragraph{Preprocessing} The FETI preprocessing also contains a parallel loop, where for each subdomain, we first perform the numerical factorization and copy the factors' non-zero values to the GPU. Then, we allocate the workspace buffers and all temporary matrices in the temporary memory allocator. Next, we convert the gluing matrices $\tilde{\mB}_i$ to dense on the GPU and submit the TRSM and matrix multiplication kernels that perform the actual assembly of the local dual operator. Finally, the temporarily allocated matrices are deallocated, which concludes the parallel loop. Because all the operations submitted to the GPU are asynchronous, we need to perform a synchronization to wait for all the GPU work to finish.

The asynchronicity of the GPU operations is important. While the GPU is executing kernels corresponding to one of the previous subdomains, we can perform numerical factorization and submit kernels corresponding to another subdomain in the following iteration of the subdomain loop. After the first iteration, we, therefore, achieve CPU-GPU computation overlap. Furthermore, since we use multiple CUDA streams, kernels and memory transfers coming from different subdomains can run concurrently, thus increasing GPU occupancy and allowing copy-compute overlap.

\paragraph{Application} When we need to apply the dual operator, we first scatter the cluster-wide dual vector to the subdomains. Then, we submit all the GEMV or SYMV\footnote{If $\tilde{\mF}_i$ is symmetric, we only store its upper or lower triangle. We share memory such that two opposite triangles are stored in a single allocation.} kernels to perform the matrix-vector multiplications $\tilde{\vq}_i = \tilde{\mF}_i \tilde{\vp}_i$ for each subdomain in the cluster. After this, we gather the results back into the cluster-wide dual vector. Because all the operations are again asynchronous, the function ends with a synchronization that waits for all the GPU operations to finish.

The scatter and gather can be performed on the CPU or on the GPU, the required dual vectors are copied to the GPU at the appropriate time. The behavior is controlled by a parameter, see Section~\ref{sec:settings_description} for more details.

\subsection{Parameters of the explicit assembly process}
\label{sec:settings_description}

The most time-consuming kernel during the explicit assembly of $\tilde{\mF}_i$ is TRSM.
This kernel is usually provided by both sparse and dense BLAS libraries, i.e., one can call it with factors in sparse or dense format.
Both versions also allow us to provide the matrices (factor and right-hand side) in row-major or column-major order (indirectly through other parameters).
These options have an impact on the performance and memory demands of the TRSM kernel and the whole assembly process.
Moreover, in the general form, TRSM is called twice (forward and backward solve). For SPD systems with $\mL \mL^\top$ or $\mU^\top \mU$ factorization, it is possible to avoid the second TRSM using the SYRK function (as described below). 

\begin{table}
    \centering
    \caption{Overview of parameters for the explicit assembly of $\tilde{\mF}_i$ using GPUs.}
    \label{tab:settings_overview}
    \begin{tabular}{ll}
        \toprule
        Setting & Options \\
        \midrule
        Path & TRSM, SYRK \\
        Forward solve factor storage & sparse, dense \\
        Backward solve factor storage & sparse, dense \\
        \midrule
        Forward solve factor order & row-major, column-major \\
        Backward solve factor order & row-major, column-major \\
        RHS memory order & row-major, column-major \\
        \midrule
        Scatter and gather & CPU, GPU \\
        \bottomrule
    \end{tabular}
\end{table}

All the parameters are summarized in Table~\ref{tab:settings_overview}. We follow with a more detailed explanation of their meaning.

\textbf{Path} determines the matrix operations that are performed to assemble $\tilde{\mF}_i$ in case of SPD systems. There are two possibilities named after the second invoked kernel:
\begin{itemize}
\item TRSM: $\tilde{\mF}_i = \tilde{\mB}_i (\mU_i^{-1} (\mU_i^{-\top} \tilde{\mB}_i^\top))$.
\item SYRK: $\tilde{\mF}_i = (\mU_i^{-\top} \tilde{\mB}_i^\top)^\top (\mU_i^{-\top} \tilde{\mB}_i^\top)$.
\end{itemize}
For the TRSM version, two triangular solves are followed by a sparse-dense matrix-matrix multiplication. In SYRK, only the first triangular solve is performed, after which we do a dense matrix-matrix multiplication -- SYRK.
Note that the performance difference is practically only in replacing the TRSM kernel with SYRK, avoiding the SPMM has a negligible effect.
In the case of dense factor storage, SYRK is theoretically faster -- both kernels are level 3 BLAS, but we call SYRK with smaller matrices. With sparse factors, the theoretically better option depends on the sparsity of the factors, and we leave the decision up to experiments.

\textbf{Factor storage} denotes if we use the sparse TRSM from cuSPARSE or the dense TRSM from cuBLAS.
Due to the higher density of the factors caused by fill-in, using the dense TRSM might be faster, even though it cannot utilize the sparsity.
The sparse-to-dense conversion is performed on the GPU to minimize the amount of data transferred.

\textbf{Factor order} denotes the memory order of the factor passed to the TRSM kernel. For sparse TRSM, this parameter controls whether CSR or CSC format is used for the factor. The GPU library functions typically assume column-major order of dense matrices or the CSR format for sparse, but we can tweak other parameters to indirectly enable this option. If necessary, we convert the matrix order (which is equivalent to transposing the matrix) on the CPU.

\textbf{RHS memory order} specifies the memory order of dense RHS (right-hand side) and solution matrices passed to/from the TRSM routine.

\textbf{Scatter and gather} can be performed either on the CPU or on the GPU. If the CPU is used, then the subdomain-wide dual vectors are individually copied to and from the GPU right before and after the GEMV or SYMV kernel is submitted. If we use the GPU, we copy the whole cluster-wide dual vector to the GPU, submit a single kernel that performs its scattering, and submit all the GEMV/SYMV kernels. Finally, we perform the gather and copy the output dual vector back to the CPU. The difference between the options is that the CPU version submits more GPU operations, which causes more overheads, but allows for more concurrency and copy-compute overlap.

\section{Results}
\label{sec:performance}

In this section, the performance of the accelerated FETI solver is presented.
The comparison is divided into three sections: A) choosing the optimal parameters for the explicit assembly of $\tilde{\mF}_i$ and its application, B) comparison of different versions of the dual operator, and C) the overall performance of the FETI solver.

The measurements were performed for problems of heat transfer and linear elasticity, both in 2D and 3D, and with varying sizes of subdomains (denoted as degrees of freedom, DOFs, in the graphs). With decreasing the subdomain size, we increase their count (up to 2000) to have the total number of DOFs approximately constant, mainly to keep the runtime at a reasonable scale\footnote{Note that all GPU operations bring some overhead, and for small subdomains, where the operations themselves are quick, the overheads can dominate the runtime.}. That was around 8.4M DOFs for 2D and 1.1M DOFs for 3D problems.
We used a square or cube domain discretized into a mesh composed of triangles or tetrahedral elements.

The performance was tested on the GPU partition of the Karolina cluster at IT4Innovations~\cite{karolinadocs}.
The partition is equipped with two 64-core AMD EPYC 7763 processors with 1 TB of DDR4 memory and has 8 NVIDIA A100 GPUs with 40 GB of HBM2 memory.
For most of the experiments, we use only a single GPU and a proportional fraction of CPU cores and memory, that is, 16 cores and 128 GB of memory (which corresponds to a single NUMA domain).
This configuration reflects production runs with fully utilized nodes since using a single MPI process per NUMA domain with a single GPU usually delivers optimal performance.
GPU kernels were submitted using 16 CUDA streams, i.e., one stream per OpenMP thread.
The source code and experiment scripts are a part of the ESPRESO library and can be found in~\cite{espresogithub}. All the measured data are available at~\cite{experimentdatasetfigshare}.

We tested two versions of CUDA libraries, \textit{legacy} and \textit{modern}. The legacy stands for CUDA toolkit 11.7, where the legacy cuSPARSE API was used. The modern stands for CUDA toolkit 12.4, where we use the generic cuSPARSE API functions.
On the CPU, we used two sparse linear solver libraries -- CHOLMOD~\cite{cholmod} from SuiteSparse 7.6.0 and PARDISO from Intel MKL 2024.2.0~\cite{intelmkl}. Both libraries used Metis to reduce fill-in.
MKL PARDISO provides the augmented incomplete factorization algorithm for the assembly of explicit $\tilde{\mF}_i$ via Schur complement on the CPU. It does not allow the extraction of the factors $\mL_i$ and $\mU_i$, so it cannot be used for GPU acceleration of the explicit assembly.
CHOLMOD provides functions for the extraction of factors from the solver.

\subsection{Optimal parameters of the assembly}

As we discussed in Section~\ref{sec:settings_description}, the explicit assembly function has several parameters that influence both runtime and memory requirements. Now, we show the optimal values of those parameters according to our experiments.
To account for potential dependencies, we tested all possible combinations of the parameters.
Listing and reasoning through all of the results would not fit in this paper; therefore, in this text, we provide only a representative example for each parameter.

\begin{table}
    \centering
    \caption{Optimal parameters for the explicit assembly of the local dual operator.}
    \label{tab:optimal_settings}
    \begin{tabular}{lll}
    \toprule
    CUDA library                          & legacy (v11.7)                 & modern (v12.4)               \\
    \midrule
    path                                  & SYRK                           & SYRK                         \\
    \midrule
    \multirow{3}{2.5cm}{factor storage}   & 2D: sparse                     & \multirow{3}{2cm}{dense}     \\
                                          & 3D: $<$ 12k DOFs: dense        &                              \\
                                          & 3D: $>$ 12k DOFs: sparse       &                              \\
    \midrule
    \multirow{2}{2.5cm}{factor order}     & sparse: row-major              & \multirow{2}{2cm}{col-major} \\
                                          & dense: col-major               &                              \\
    \midrule
    \multirow{2}{2.5cm}{RHS memory order} & \multirow{2}{2.5cm}{row-major} & 2D: col-major                \\
                                          &                                & 3D: row-major                \\
    \bottomrule
    \end{tabular}
\end{table}

An overview of the optimal parameter settings can be found in Table~\ref{tab:optimal_settings}.
The configuration for modern CUDA is straightforward; for legacy CUDA, the settings are more problem-dependent, but the explicit assembly process is faster there. In our implementation, we have an option to auto-configure these parameters based on the problem that is being solved. In the following paragraphs, we will describe the reasoning behind the optimal parameter settings in more detail.

\begin{figure}
 \centering
 \input{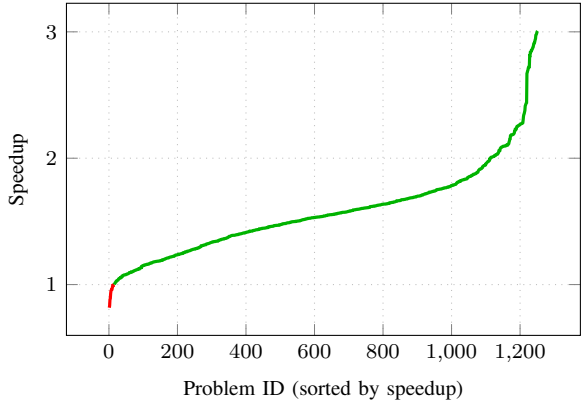}
 \caption{Speedup of SYRK path compared to TRSM in the explicit assembly of $\tilde{\mF}_i$.}
 \label{fig:settings_path}
\end{figure}

\paragraph{Path} For the path parameter, it was faster to use SYRK for the majority of the problems.
Fig.~\ref{fig:settings_path} shows the FETI preprocessing speedup of the SYRK path compared to TRSM for all tested configurations using both CUDA versions.
The average speedup was 1.58. TRSM was better only in 16 cases for very small subdomains.
In some cases, TRSM failed because of excessive memory consumption or because it exceeded our time limit.
This experiment confirms the better performance of SYRK for dense factors.
In the case of sparse factors, the SYRK is still better due to the relatively high fill-in of the factors.

\begin{figure}
 \centering
 \begin{tikzpicture}
  \begin{loglogaxis}[
    width=.95\columnwidth, 
    height=6cm,
    log basis x=2,
    log basis y=10,
    xlabel={Number of DOFs per subdomain},
    ylabel={Time per subdomain [ms]},
    xmajorgrids=true,
    ymajorgrids=true,
    every axis plot/.append style={line width=1.2pt},
    grid style=dotted,
    xmin=242,
    xmax=97469,
    ymin=94,
    ymax=50624,
    label style={font=\footnotesize},
    tick label style={font=\footnotesize},
    legend style={font=\footnotesize, draw=none, at={(0.95,0.05)}, anchor=south east, legend columns=1},
    legend cell align={left}
  ]
    \addplot[dashed, color={blue}, mark=none] coordinates {
      (  343 ,  921.132 )
      (  729 , 2950.456 )
      ( 1331 , 8080.394 )
      ( 2197 , 8767.602 )
    };
    \addplot[solid, color={blue}, mark=none] coordinates {
      (   343 ,   153.017 )
      (   729 ,   337.743 )
      (  1331 ,   820.683 )
      (  2197 ,  1055.936 )
      (  4913 ,  2862.325 )
      (  9261 ,  5725.716 )
      ( 19683 , 15992.118 )
      ( 35937 , 36759.575 )
    };
    \addplot[dashed, color={red}, mark=none] coordinates {
      (   343 ,   133.735 )
      (   729 ,   351.376 )
      (  1331 ,   964.267 )
      (  2197 ,  1163.215 )
      (  4913 ,  2727.881 )
      (  9261 ,  4686.846 )
      ( 19683 , 10529.750 )
      ( 35937 , 18646.731 )
      ( 68921 , 30323.116 )
    };
    \addplot[solid, color={red}, mark=none] coordinates {
      (   343 ,   151.081 )
      (   729 ,   338.697 )
      (  1331 ,   757.922 )
      (  2197 ,   984.992 )
      (  4913 ,  2705.722 )
      (  9261 ,  5504.220 )
      ( 19683 , 15722.003 )
      ( 35937 , 35797.209 )
    };
    \addlegendentry{sparse, modern}
    \addlegendentry{dense, modern}
    \addlegendentry{sparse, legacy}
    \addlegendentry{dense, legacy}
  \end{loglogaxis}
\end{tikzpicture}
 \caption{Comparison of factor storage in explicit assembly of $\tilde{\mF}_i$.
 Settings: heat transfer 3D, quadratic tetrahedra, SYRK path.}
 \label{fig:settings_factor_storage}
\end{figure}
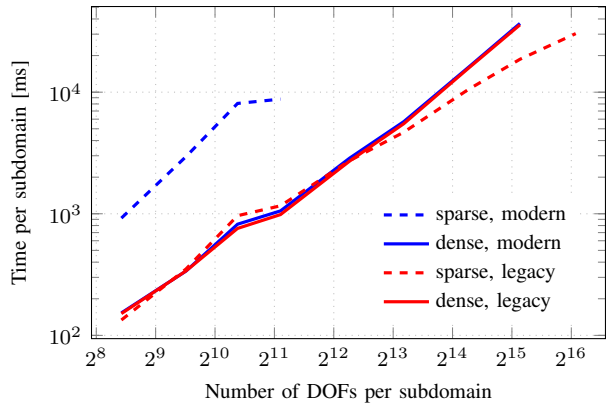

\paragraph{Factor storage} Fig.~\ref{fig:settings_factor_storage} shows the comparison of factor storage for both versions of CUDA for a 3D heat transfer problem.
The optimal setting is highly dependent on the problem dimensionality, subdomain size, and even the used version of the cuSPARSE API; however, the behavior between forward and backward TRSM is almost identical.

The modern CUDA with generic cuSPARSE API has a very underperforming sparse TRSM kernel, as one can observe in the graph. The kernel also requires very large persistently allocated memory buffers, which very significantly limits the maximum problem size we are able to handle. It is, therefore, always better to use dense TRSM across all other parameters, problem types, and subdomain sizes with modern CUDA. 

The TRSM from legacy cuSPARSE performs much better, mainly due to the used block algorithm.
For 2D meshes, the factors are very sparsely populated, and it is better to use them as sparse.
As can be observed in the graph, for 3D meshes where the factors are denser, it is unclear which is the better option.
For large subdomains that contain more than 12000 DOFs, the sparsity is still high enough to make sparse storage better.
However, for subdomains smaller than that, the factors are denser, and the choice is challenging.
We decided to use dense storage there because it was the better option for 51/64 measurements of medium-sized subdomains (1000-12000 DOFs).
However, for a production run, we recommend performing a few benchmarks to select the right factor storage for the specific problem that is being solved.

\paragraph{Factor order} This parameter influences the runtime of the assembly process only very little (hence, we do not provide any graph). The main difference is in the size of the workspace buffers in the sparse TRSM from legacy CUDA. If a column-major (CSC) factor is used, the function requires additional memory (both persistent and temporary) with the size around the size of the factor, and it is, therefore, better to use row-major order there. In modern CUDA, the workspace buffer size is not affected by any parameters. For dense matrices, we use column-major order.

\paragraph{RHS memory order} For dense factor storage and 2D subdomains, where the number of columns in the RHS is proportionally small, column-major order was slightly faster. For dense storage and 3D subdomains, where the right-hand side is proportionally wider, row-major order was better.

For sparse factor storage, workspace buffer sizes again play a major role. With column-major right-hand side, the sparse TRSM kernel in legacy CUDA required an additional temporary workspace memory equal to the size of the right-hand side matrix, most probably for its transpose (row-major copy). Therefore, with sparse factors, we use row-major right-hand side and solution.

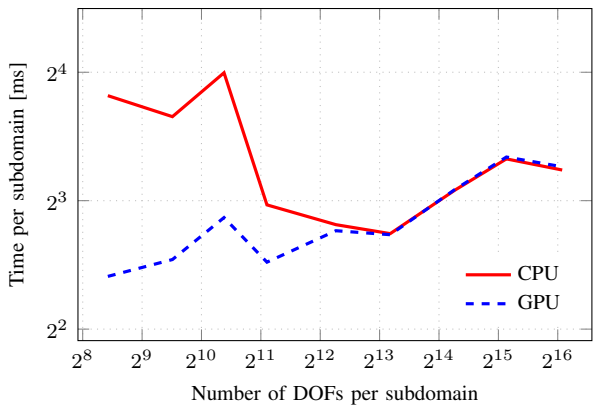
\begin{figure}
 \centering
 \begin{tikzpicture}
  \begin{loglogaxis}[
    width=.95\columnwidth, 
    height=6cm,
    log basis x=2,
    log basis y=2,
    xlabel={Number of DOFs per subdomain},
    ylabel={Time per subdomain [ms]},
    xmajorgrids=true,
    ymajorgrids=true,
    every axis plot/.append style={line width=1.2pt},
    grid style=dotted,
    xmin=242,
    xmax=97469,
    ymin=3.7587,
    ymax=22.556,
    label style={font=\footnotesize},
    tick label style={font=\footnotesize},
    legend style={font=\footnotesize, draw=none, at={(0.85,0.05)}, anchor=south, legend columns=1},
    legend cell align={left}
  ]
    \addplot[solid, color={red}, mark=none] coordinates {
      (   343 , 14.095392 )
      (   729 , 12.586217 )
      (  1331 , 15.949619 )
      (  2197 ,  7.819464 )
      (  4913 ,  7.027349 )
      (  9261 ,  6.695428 )
      ( 19683 ,  8.463807 )
      ( 35937 , 10.020573 )
      ( 68921 ,  9.436735 )
    };
    \addplot[dashed, color={blue}, mark=none] coordinates {
      (   343 ,  5.315676 )
      (   729 ,  5.821615 )
      (  1331 ,  7.301319 )
      (  2197 ,  5.739892 )
      (  4913 ,  6.801369 )
      (  9261 ,  6.655977 )
      ( 19683 ,  8.492736 )
      ( 35937 , 10.121843 )
      ( 68921 ,  9.604714 )
    };
    \addlegendentry{CPU}
    \addlegendentry{GPU}
  \end{loglogaxis}
\end{tikzpicture}
 \caption{Comparison of performing scatter and gather on CPU or GPU. Settings: heat transfer 3D, quadratic tetrahedra.}
 \label{fig:settings_scatter_gather}
\end{figure}

\paragraph{Scatter and gather} Fig.~\ref{fig:settings_scatter_gather} shows the per-subdomain application time when the scatter and gather operations are performed on the CPU and on the GPU.
For small subdomains, using the CPU is slower because of the overhead of submitting more GPU operations.
With increasing the subdomain size, the overheads become less dominant, and for very large subdomains, using the CPU is slightly faster because it allows for more concurrency. However, we will use the GPU because it is better for a wider range of subdomain sizes, and the difference for the large subdomains was only 3 \% on average.

\subsection{Comparison of preprocessing and application performance}

This section compares all available approaches for assembling and applying the dual operator.
All approaches are summarized in Table~\ref{tab:dual_approaches}.
Apart from the explicit GPU approach, we also implemented and tested the implicit GPU approach, explicit and implicit CPU approaches, and a hybrid approach equivalent to the original acceleration attempts, as seen in~\cite{BDDS_ACC,ESPRESO-SC}.
For the implicit approaches, only the factorization of matrices $\mK_i$ was performed during the FETI preprocessing. In the implicit GPU approach, the factors were also copied to GPU.
In the case of explicit approaches, after factorization of matrices $\mK_i$, $\tilde{\mF}_i$ were explicitly assembled on an architecture defined in the description.

\begin{table}
    \centering
    \caption{A description of all tested approaches for the dual operator.}
    \label{tab:dual_approaches}
    \begin{tabular}{ll}
    \toprule
    approach      & description           \\
    \midrule
    impl\_mkl     & the MKL PARDISO solver on CPU \\
    impl\_cholmod & the CHOLMOD solver on CPU\\
    impl\_legacy  & CUDA legacy with factors from CHOLMOD \\
    impl\_modern  & CUDA modern with factors from CHOLMOD \\
    \midrule
    expl\_mkl     & aug. incomplete fact. from MKL PARDISO on CPU \\
    expl\_cholmod & TRSM with the CHOLMOD solver on CPU \\
    expl\_legacy  & CUDA legacy with factors from CHOLMOD \\
    expl\_modern  & CUDA modern with factors from CHOLMOD \\
    \midrule
    expl\_hybrid  & assembly expl\_mkl, application CUDA modern \\
    \bottomrule
    \end{tabular}
\end{table}

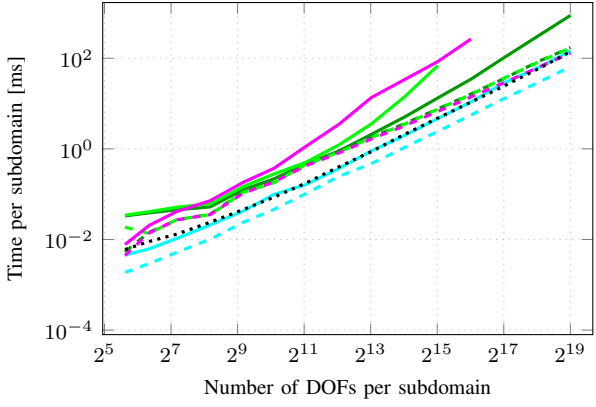
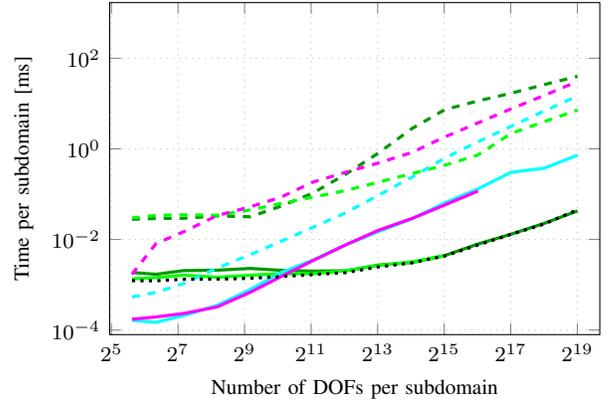
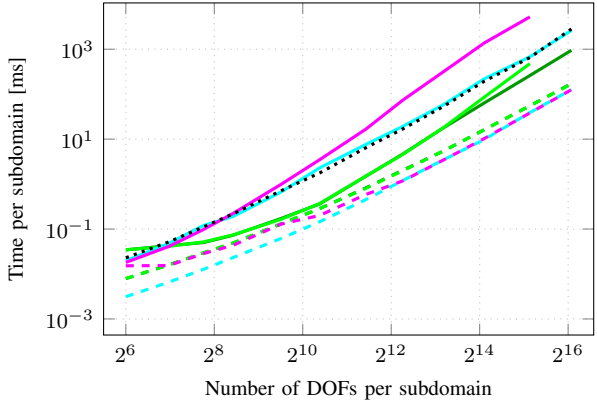
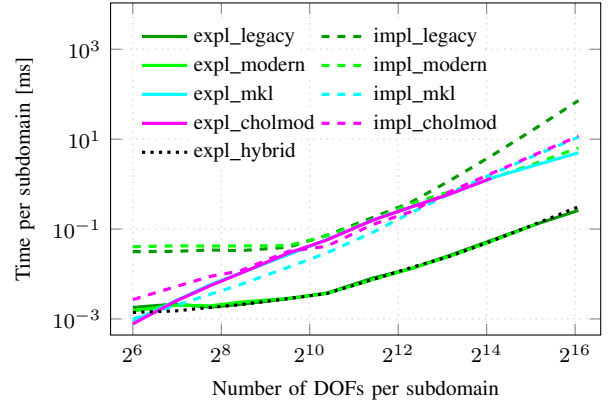
\begin{figure*}
    \begin{subfigure}[b]{0.47\textwidth}
        \centering
        \begin{tikzpicture}
  \begin{loglogaxis}[
    width=.95\columnwidth, 
    height=6cm,
    log basis x=2,
    log basis y=10,
    xlabel={Number of DOFs per subdomain},
    ylabel={Time per subdomain [ms]},
    xmajorgrids=true,
    ymajorgrids=true,
    every axis plot/.append style={line width=1.2pt},
    grid style=dotted,
    xmin=3.081e+01,
    xmax=8.360e+05,
    ymin=7.945e-05,
    ymax=1.696e+03,
    label style={font=\footnotesize},
    tick label style={font=\footnotesize},
    legend style={font=\footnotesize, draw=none, at={(0.5,-0.1)}, anchor=north},
    legend cell align={left}
  ]
    \addplot[solid, color={green!60!black}, mark=none] coordinates {
      ( 4.900e+01 , 3.309e-02 )
      ( 8.100e+01 , 3.871e-02 )
      ( 1.440e+02 , 4.509e-02 )
      ( 2.890e+02 , 5.244e-02 )
      ( 5.760e+02 , 1.256e-01 )
      ( 1.089e+03 , 2.170e-01 )
      ( 2.116e+03 , 4.672e-01 )
      ( 4.225e+03 , 8.953e-01 )
      ( 8.281e+03 , 2.082e+00 )
      ( 1.664e+04 , 5.077e+00 )
      ( 3.312e+04 , 1.331e+01 )
      ( 6.605e+04 , 3.456e+01 )
      ( 1.318e+05 , 1.033e+02 )
      ( 2.632e+05 , 3.029e+02 )
      ( 5.256e+05 , 8.828e+02 )
    };
    \addplot[dashed, color={green!60!black}, mark=none] coordinates {
      ( 4.900e+01 , 5.361e-03 )
      ( 8.100e+01 , 1.443e-02 )
      ( 1.440e+02 , 2.681e-02 )
      ( 2.890e+02 , 3.599e-02 )
      ( 5.760e+02 , 1.137e-01 )
      ( 1.089e+03 , 1.926e-01 )
      ( 2.116e+03 , 4.276e-01 )
      ( 4.225e+03 , 9.118e-01 )
      ( 8.281e+03 , 1.810e+00 )
      ( 1.664e+04 , 3.620e+00 )
      ( 3.312e+04 , 7.619e+00 )
      ( 6.605e+04 , 1.625e+01 )
      ( 1.318e+05 , 3.566e+01 )
      ( 2.632e+05 , 8.274e+01 )
      ( 5.256e+05 , 1.727e+02 )
    };
    \addplot[solid, color=green, mark=none] coordinates {
      ( 4.900e+01 , 3.445e-02 )
      ( 8.100e+01 , 4.101e-02 )
      ( 1.440e+02 , 5.157e-02 )
      ( 2.890e+02 , 6.147e-02 )
      ( 5.760e+02 , 1.478e-01 )
      ( 1.089e+03 , 2.787e-01 )
      ( 2.116e+03 , 5.051e-01 )
      ( 4.225e+03 , 1.224e+00 )
      ( 8.281e+03 , 3.556e+00 )
      ( 1.664e+04 , 1.427e+01 )
      ( 3.312e+04 , 6.890e+01 )
      ( 6.605e+04 , nan )
      ( 1.318e+05 , nan )
      ( 2.632e+05 , nan )
      ( 5.256e+05 , nan )
    };
    \addplot[dashed, color=green, mark=none] coordinates {
      ( 4.900e+01 , 1.872e-02 )
      ( 8.100e+01 , 1.354e-02 )
      ( 1.440e+02 , 2.764e-02 )
      ( 2.890e+02 , 3.543e-02 )
      ( 5.760e+02 , 1.105e-01 )
      ( 1.089e+03 , 1.785e-01 )
      ( 2.116e+03 , 4.576e-01 )
      ( 4.225e+03 , 8.475e-01 )
      ( 8.281e+03 , 1.771e+00 )
      ( 1.664e+04 , 3.518e+00 )
      ( 3.312e+04 , 7.526e+00 )
      ( 6.605e+04 , 1.590e+01 )
      ( 1.318e+05 , 3.657e+01 )
      ( 2.632e+05 , 8.035e+01 )
      ( 5.256e+05 , 1.709e+02 )
    };
    \addplot[solid, color={rgb,255:red,0;green,255;blue,255}, mark=none] coordinates {
      ( 4.900e+01 , 4.521e-03 )
      ( 8.100e+01 , 6.204e-03 )
      ( 1.440e+02 , 1.055e-02 )
      ( 2.890e+02 , 2.107e-02 )
      ( 5.760e+02 , 4.256e-02 )
      ( 1.089e+03 , 9.878e-02 )
      ( 2.116e+03 , 1.625e-01 )
      ( 4.225e+03 , 3.680e-01 )
      ( 8.281e+03 , 8.985e-01 )
      ( 1.664e+04 , 1.986e+00 )
      ( 3.312e+04 , 4.660e+00 )
      ( 6.605e+04 , 1.090e+01 )
      ( 1.318e+05 , 2.872e+01 )
      ( 2.632e+05 , 6.112e+01 )
      ( 5.256e+05 , 1.437e+02 )
    };
    \addplot[dashed, color={rgb,255:red,0;green,255;blue,255}, mark=none] coordinates {
      ( 4.900e+01 , 1.875e-03 )
      ( 8.100e+01 , 2.946e-03 )
      ( 1.440e+02 , 5.190e-03 )
      ( 2.890e+02 , 1.014e-02 )
      ( 5.760e+02 , 2.398e-02 )
      ( 1.089e+03 , 4.597e-02 )
      ( 2.116e+03 , 1.026e-01 )
      ( 4.225e+03 , 2.479e-01 )
      ( 8.281e+03 , 4.700e-01 )
      ( 1.664e+04 , 1.081e+00 )
      ( 3.312e+04 , 2.431e+00 )
      ( 6.605e+04 , 5.583e+00 )
      ( 1.318e+05 , 1.293e+01 )
      ( 2.632e+05 , 2.866e+01 )
      ( 5.256e+05 , 6.534e+01 )
    };
    \addplot[solid, color={rgb,255:red,255;green,0;blue,255}, mark=none] coordinates {
      ( 4.900e+01 , 7.749e-03 )
      ( 8.100e+01 , 2.041e-02 )
      ( 1.440e+02 , 4.143e-02 )
      ( 2.890e+02 , 7.179e-02 )
      ( 5.760e+02 , 1.827e-01 )
      ( 1.089e+03 , 3.711e-01 )
      ( 2.116e+03 , 1.131e+00 )
      ( 4.225e+03 , 3.490e+00 )
      ( 8.281e+03 , 1.355e+01 )
      ( 1.664e+04 , 3.401e+01 )
      ( 3.312e+04 , 8.469e+01 )
      ( 6.605e+04 , 2.656e+02 )
      ( 1.318e+05 , nan )
      ( 2.632e+05 , nan )
      ( 5.256e+05 , nan )
    };
    \addplot[dashed, color={rgb,255:red,255;green,0;blue,255}, mark=none] coordinates {
      ( 4.900e+01 , 4.427e-03 )
      ( 8.100e+01 , 1.391e-02 )
      ( 1.440e+02 , 2.718e-02 )
      ( 2.890e+02 , 3.538e-02 )
      ( 5.760e+02 , 1.055e-01 )
      ( 1.089e+03 , 1.854e-01 )
      ( 2.116e+03 , 4.236e-01 )
      ( 4.225e+03 , 8.154e-01 )
      ( 8.281e+03 , 1.654e+00 )
      ( 1.664e+04 , 3.250e+00 )
      ( 3.312e+04 , 6.703e+00 )
      ( 6.605e+04 , 1.400e+01 )
      ( 1.318e+05 , 2.892e+01 )
      ( 2.632e+05 , 6.122e+01 )
      ( 5.256e+05 , 1.302e+02 )
    };
    \addplot[dotted, color=black, mark=none] coordinates {
      ( 4.900e+01 , 6.003e-03 )
      ( 8.100e+01 , 9.027e-03 )
      ( 1.440e+02 , 1.310e-02 )
      ( 2.890e+02 , 2.401e-02 )
      ( 5.760e+02 , 4.569e-02 )
      ( 1.089e+03 , 8.575e-02 )
      ( 2.116e+03 , 1.755e-01 )
      ( 4.225e+03 , 4.016e-01 )
      ( 8.281e+03 , 8.678e-01 )
      ( 1.664e+04 , 2.101e+00 )
      ( 3.312e+04 , 4.768e+00 )
      ( 6.605e+04 , 1.074e+01 )
      ( 1.318e+05 , 2.444e+01 )
      ( 2.632e+05 , 5.708e+01 )
      ( 5.256e+05 , 1.395e+02 )
    };
  \end{loglogaxis}
\end{tikzpicture}
        \caption{Heat transfer 2D, preprocessing}
        \label{fig:time_karolina_2d_update}
    \end{subfigure}
    \hfill
    \begin{subfigure}[b]{0.47\textwidth}
        \centering
        \begin{tikzpicture}
  \begin{loglogaxis}[
    width=.95\columnwidth, 
    height=6cm,
    log basis x=2,
    log basis y=10,
    xlabel={Number of DOFs per subdomain},
    ylabel={Time per subdomain [ms]},
    xmajorgrids=true,
    ymajorgrids=true,
    every axis plot/.append style={line width=1.2pt},
    grid style=dotted,
    xmin=3.081e+01,
    xmax=8.360e+05,
    ymin=7.945e-05,
    ymax=1.696e+03,
    label style={font=\footnotesize},
    tick label style={font=\footnotesize},
    legend style={font=\footnotesize, draw=none, at={(0.5,-0.1)}, anchor=north},
    legend cell align={left}
  ]
    \addplot[solid, color={green!60!black}, mark=none] coordinates {
      ( 4.900e+01 , 1.870e-03 )
      ( 8.100e+01 , 1.706e-03 )
      ( 1.440e+02 , 2.060e-03 )
      ( 2.890e+02 , 2.098e-03 )
      ( 5.760e+02 , 2.298e-03 )
      ( 1.089e+03 , 2.086e-03 )
      ( 2.116e+03 , 2.001e-03 )
      ( 4.225e+03 , 2.040e-03 )
      ( 8.281e+03 , 2.728e-03 )
      ( 1.664e+04 , 3.100e-03 )
      ( 3.312e+04 , 4.431e-03 )
      ( 6.605e+04 , 7.940e-03 )
      ( 1.318e+05 , 1.309e-02 )
      ( 2.632e+05 , 2.276e-02 )
      ( 5.256e+05 , 4.272e-02 )
    };
    \addplot[dashed, color={green!60!black}, mark=none] coordinates {
      ( 4.900e+01 , 2.788e-02 )
      ( 8.100e+01 , 2.909e-02 )
      ( 1.440e+02 , 2.937e-02 )
      ( 2.890e+02 , 3.312e-02 )
      ( 5.760e+02 , 3.168e-02 )
      ( 1.089e+03 , 5.415e-02 )
      ( 2.116e+03 , 1.053e-01 )
      ( 4.225e+03 , 2.778e-01 )
      ( 8.281e+03 , 8.029e-01 )
      ( 1.664e+04 , 2.792e+00 )
      ( 3.312e+04 , 7.341e+00 )
      ( 6.605e+04 , 1.144e+01 )
      ( 1.318e+05 , 1.688e+01 )
      ( 2.632e+05 , 2.623e+01 )
      ( 5.256e+05 , 3.992e+01 )
    };
    \addplot[solid, color=green, mark=none] coordinates {
      ( 4.900e+01 , 1.351e-03 )
      ( 8.100e+01 , 1.444e-03 )
      ( 1.440e+02 , 1.630e-03 )
      ( 2.890e+02 , 1.470e-03 )
      ( 5.760e+02 , 1.628e-03 )
      ( 1.089e+03 , 1.758e-03 )
      ( 2.116e+03 , 1.751e-03 )
      ( 4.225e+03 , 2.049e-03 )
      ( 8.281e+03 , 2.655e-03 )
      ( 1.664e+04 , 3.196e-03 )
      ( 3.312e+04 , 4.407e-03 )
      ( 6.605e+04 , nan )
      ( 1.318e+05 , nan )
      ( 2.632e+05 , nan )
      ( 5.256e+05 , nan )
    };
    \addplot[dashed, color=green, mark=none] coordinates {
      ( 4.900e+01 , 2.990e-02 )
      ( 8.100e+01 , 3.411e-02 )
      ( 1.440e+02 , 3.510e-02 )
      ( 2.890e+02 , 3.416e-02 )
      ( 5.760e+02 , 4.497e-02 )
      ( 1.089e+03 , 6.207e-02 )
      ( 2.116e+03 , 8.580e-02 )
      ( 4.225e+03 , 1.209e-01 )
      ( 8.281e+03 , 1.838e-01 )
      ( 1.664e+04 , 2.834e-01 )
      ( 3.312e+04 , 4.350e-01 )
      ( 6.605e+04 , 7.315e-01 )
      ( 1.318e+05 , 2.193e+00 )
      ( 2.632e+05 , 3.985e+00 )
      ( 5.256e+05 , 7.215e+00 )
    };
    \addplot[solid, color={rgb,255:red,0;green,255;blue,255}, mark=none] coordinates {
      ( 4.900e+01 , 1.641e-04 )
      ( 8.100e+01 , 1.484e-04 )
      ( 1.440e+02 , 2.092e-04 )
      ( 2.890e+02 , 3.520e-04 )
      ( 5.760e+02 , 7.869e-04 )
      ( 1.089e+03 , 1.865e-03 )
      ( 2.116e+03 , 3.486e-03 )
      ( 4.225e+03 , 7.655e-03 )
      ( 8.281e+03 , 1.489e-02 )
      ( 1.664e+04 , 2.847e-02 )
      ( 3.312e+04 , 6.407e-02 )
      ( 6.605e+04 , 1.310e-01 )
      ( 1.318e+05 , 3.037e-01 )
      ( 2.632e+05 , 3.738e-01 )
      ( 5.256e+05 , 7.253e-01 )
    };
    \addplot[dashed, color={rgb,255:red,0;green,255;blue,255}, mark=none] coordinates {
      ( 4.900e+01 , 5.373e-04 )
      ( 8.100e+01 , 6.813e-04 )
      ( 1.440e+02 , 1.067e-03 )
      ( 2.890e+02 , 2.340e-03 )
      ( 5.760e+02 , 4.628e-03 )
      ( 1.089e+03 , 9.099e-03 )
      ( 2.116e+03 , 1.846e-02 )
      ( 4.225e+03 , 3.910e-02 )
      ( 8.281e+03 , 9.390e-02 )
      ( 1.664e+04 , 2.376e-01 )
      ( 3.312e+04 , 6.243e-01 )
      ( 6.605e+04 , 1.434e+00 )
      ( 1.318e+05 , 3.143e+00 )
      ( 2.632e+05 , 6.862e+00 )
      ( 5.256e+05 , 1.467e+01 )
    };
    \addplot[solid, color={rgb,255:red,255;green,0;blue,255}, mark=none] coordinates {
      ( 4.900e+01 , 1.739e-04 )
      ( 8.100e+01 , 1.950e-04 )
      ( 1.440e+02 , 2.333e-04 )
      ( 2.890e+02 , 3.247e-04 )
      ( 5.760e+02 , 6.840e-04 )
      ( 1.089e+03 , 1.510e-03 )
      ( 2.116e+03 , 3.408e-03 )
      ( 4.225e+03 , 7.646e-03 )
      ( 8.281e+03 , 1.595e-02 )
      ( 1.664e+04 , 2.890e-02 )
      ( 3.312e+04 , 5.780e-02 )
      ( 6.605e+04 , 1.170e-01 )
      ( 1.318e+05 , nan )
      ( 2.632e+05 , nan )
      ( 5.256e+05 , nan )
    };
    \addplot[dashed, color={rgb,255:red,255;green,0;blue,255}, mark=none] coordinates {
      ( 4.900e+01 , 1.696e-03 )
      ( 8.100e+01 , 8.152e-03 )
      ( 1.440e+02 , 1.484e-02 )
      ( 2.890e+02 , 3.373e-02 )
      ( 5.760e+02 , 5.270e-02 )
      ( 1.089e+03 , 8.562e-02 )
      ( 2.116e+03 , 1.841e-01 )
      ( 4.225e+03 , 3.055e-01 )
      ( 8.281e+03 , 4.861e-01 )
      ( 1.664e+04 , 8.298e-01 )
      ( 3.312e+04 , 1.832e+00 )
      ( 6.605e+04 , 3.742e+00 )
      ( 1.318e+05 , 7.636e+00 )
      ( 2.632e+05 , 1.517e+01 )
      ( 5.256e+05 , 3.078e+01 )
    };
    \addplot[dotted, color=black, mark=none] coordinates {
      ( 4.900e+01 , 1.233e-03 )
      ( 8.100e+01 , 1.228e-03 )
      ( 1.440e+02 , 1.325e-03 )
      ( 2.890e+02 , 1.331e-03 )
      ( 5.760e+02 , 1.389e-03 )
      ( 1.089e+03 , 1.519e-03 )
      ( 2.116e+03 , 1.680e-03 )
      ( 4.225e+03 , 1.859e-03 )
      ( 8.281e+03 , 2.478e-03 )
      ( 1.664e+04 , 3.054e-03 )
      ( 3.312e+04 , 4.330e-03 )
      ( 6.605e+04 , 7.561e-03 )
      ( 1.318e+05 , 1.300e-02 )
      ( 2.632e+05 , 2.196e-02 )
      ( 5.256e+05 , 4.547e-02 )
    };
  \end{loglogaxis}
\end{tikzpicture}
        \caption{Heat transfer 2D, application}
        \label{fig:time_karolina_2d_apply}
    \end{subfigure}

    \vspace{0.5cm}
    
    \begin{subfigure}[b]{0.47\textwidth}
        \centering
        \begin{tikzpicture}
  \begin{loglogaxis}[
    width=.95\columnwidth, 
    height=6cm,
    log basis x=2,
    log basis y=10,
    xlabel={Number of DOFs per subdomain},
    ylabel={Time per subdomain [ms]},
    xmajorgrids=true,
    ymajorgrids=true,
    every axis plot/.append style={line width=1.2pt},
    grid style=dotted,
    xmin=4.514e+01,
    xmax=9.771e+04,
    ymin=4.391e-04,
    ymax=1.071e+04,
    label style={font=\footnotesize},
    tick label style={font=\footnotesize},
    legend style={font=\footnotesize, draw=none, at={(0.5,-0.1)}, anchor=north},
    legend cell align={left}
  ]
    \addplot[solid, color={green!60!black}, mark=none] coordinates {
      ( 6.400e+01 , 3.440e-02 )
      ( 1.250e+02 , 4.276e-02 )
      ( 2.160e+02 , 5.005e-02 )
      ( 3.430e+02 , 7.273e-02 )
      ( 7.290e+02 , 1.719e-01 )
      ( 1.331e+03 , 3.599e-01 )
      ( 2.744e+03 , 1.488e+00 )
      ( 4.913e+03 , 4.671e+00 )
      ( 9.261e+03 , 1.773e+01 )
      ( 1.758e+04 , 6.374e+01 )
      ( 3.594e+04 , 2.597e+02 )
      ( 6.892e+04 , 9.475e+02 )
    };
    \addplot[dashed, color={green!60!black}, mark=none] coordinates {
      ( 6.400e+01 , 7.893e-03 )
      ( 1.250e+02 , 1.598e-02 )
      ( 2.160e+02 , 2.932e-02 )
      ( 3.430e+02 , 4.878e-02 )
      ( 7.290e+02 , 1.303e-01 )
      ( 1.331e+03 , 2.837e-01 )
      ( 2.744e+03 , 8.221e-01 )
      ( 4.913e+03 , 2.040e+00 )
      ( 9.261e+03 , 5.420e+00 )
      ( 1.758e+04 , 1.621e+01 )
      ( 3.594e+04 , 5.497e+01 )
      ( 6.892e+04 , 1.720e+02 )
    };
    \addplot[solid, color=green, mark=none] coordinates {
      ( 6.400e+01 , 3.463e-02 )
      ( 1.250e+02 , 4.112e-02 )
      ( 2.160e+02 , 5.173e-02 )
      ( 3.430e+02 , 7.400e-02 )
      ( 7.290e+02 , 1.647e-01 )
      ( 1.331e+03 , 3.685e-01 )
      ( 2.744e+03 , 1.495e+00 )
      ( 4.913e+03 , 4.585e+00 )
      ( 9.261e+03 , 1.797e+01 )
      ( 1.758e+04 , 8.577e+01 )
      ( 3.594e+04 , 4.778e+02 )
      ( 6.892e+04 , nan )
    };
    \addplot[dashed, color=green, mark=none] coordinates {
      ( 6.400e+01 , 7.935e-03 )
      ( 1.250e+02 , 1.574e-02 )
      ( 2.160e+02 , 2.972e-02 )
      ( 3.430e+02 , 4.882e-02 )
      ( 7.290e+02 , 1.308e-01 )
      ( 1.331e+03 , 2.862e-01 )
      ( 2.744e+03 , 8.010e-01 )
      ( 4.913e+03 , 1.983e+00 )
      ( 9.261e+03 , 5.444e+00 )
      ( 1.758e+04 , 1.613e+01 )
      ( 3.594e+04 , 5.518e+01 )
      ( 6.892e+04 , 1.681e+02 )
    };
    \addplot[solid, color={rgb,255:red,0;green,255;blue,255}, mark=none] coordinates {
      ( 6.400e+01 , 1.977e-02 )
      ( 1.250e+02 , 4.773e-02 )
      ( 2.160e+02 , 1.202e-01 )
      ( 3.430e+02 , 1.897e-01 )
      ( 7.290e+02 , 6.671e-01 )
      ( 1.331e+03 , 2.283e+00 )
      ( 2.744e+03 , 7.575e+00 )
      ( 4.913e+03 , 1.921e+01 )
      ( 9.261e+03 , 5.818e+01 )
      ( 1.758e+04 , 2.191e+02 )
      ( 3.594e+04 , 6.776e+02 )
      ( 6.892e+04 , 2.618e+03 )
    };
    \addplot[dashed, color={rgb,255:red,0;green,255;blue,255}, mark=none] coordinates {
      ( 6.400e+01 , 3.123e-03 )
      ( 1.250e+02 , 6.567e-03 )
      ( 2.160e+02 , 1.272e-02 )
      ( 3.430e+02 , 2.319e-02 )
      ( 7.290e+02 , 6.175e-02 )
      ( 1.331e+03 , 1.450e-01 )
      ( 2.744e+03 , 4.506e-01 )
      ( 4.913e+03 , 1.204e+00 )
      ( 9.261e+03 , 3.454e+00 )
      ( 1.758e+04 , 9.789e+00 )
      ( 3.594e+04 , 3.898e+01 )
      ( 6.892e+04 , 1.266e+02 )
    };
    \addplot[solid, color={rgb,255:red,255;green,0;blue,255}, mark=none] coordinates {
      ( 6.400e+01 , 1.813e-02 )
      ( 1.250e+02 , 4.074e-02 )
      ( 2.160e+02 , 9.781e-02 )
      ( 3.430e+02 , 2.218e-01 )
      ( 7.290e+02 , 9.886e-01 )
      ( 1.331e+03 , 3.473e+00 )
      ( 2.744e+03 , 1.659e+01 )
      ( 4.913e+03 , 7.391e+01 )
      ( 9.261e+03 , 3.145e+02 )
      ( 1.758e+04 , 1.377e+03 )
      ( 3.594e+04 , 5.233e+03 )
      ( 6.892e+04 , nan )
    };
    \addplot[dashed, color={rgb,255:red,255;green,0;blue,255}, mark=none] coordinates {
      ( 6.400e+01 , 1.524e-02 )
      ( 1.250e+02 , 1.537e-02 )
      ( 2.160e+02 , 2.850e-02 )
      ( 3.430e+02 , 4.341e-02 )
      ( 7.290e+02 , 1.288e-01 )
      ( 1.331e+03 , 2.005e-01 )
      ( 2.744e+03 , 5.856e-01 )
      ( 4.913e+03 , 1.181e+00 )
      ( 9.261e+03 , 3.428e+00 )
      ( 1.758e+04 , 1.029e+01 )
      ( 3.594e+04 , 3.829e+01 )
      ( 6.892e+04 , 1.257e+02 )
    };
    \addplot[dotted, color=black, mark=none] coordinates {
      ( 6.400e+01 , 2.294e-02 )
      ( 1.250e+02 , 5.143e-02 )
      ( 2.160e+02 , 1.116e-01 )
      ( 3.430e+02 , 2.169e-01 )
      ( 7.290e+02 , 7.122e-01 )
      ( 1.331e+03 , 1.763e+00 )
      ( 2.744e+03 , 6.408e+00 )
      ( 4.913e+03 , 1.690e+01 )
      ( 9.261e+03 , 5.327e+01 )
      ( 1.758e+04 , 1.938e+02 )
      ( 3.594e+04 , 6.525e+02 )
      ( 6.892e+04 , 2.819e+03 )
    };
  \end{loglogaxis}
\end{tikzpicture}
        \caption{Heat transfer 3D, preprocessing}
        \label{fig:time_karolina_3d_update}
    \end{subfigure}
    \hfill
    \begin{subfigure}[b]{0.47\textwidth}
        \centering
        \begin{tikzpicture}
  \begin{loglogaxis}[
    width=.95\columnwidth, 
    height=6cm,
    log basis x=2,
    log basis y=10,
    xlabel={Number of DOFs per subdomain},
    ylabel={Time per subdomain [ms]},
    xmajorgrids=true,
    ymajorgrids=true,
    every axis plot/.append style={line width=1.2pt},
    grid style=dotted,
    xmin=4.514e+01,
    xmax=9.771e+04,
    ymin=4.391e-04,
    ymax=1.071e+04,
    label style={font=\footnotesize},
    tick label style={font=\footnotesize},
    legend style={font=\footnotesize, draw=none, at={(0.43,0.96)}, anchor=north, legend columns=2, fill=none},
    legend cell align={left}
  ]
    \addplot[solid, color={green!60!black}, mark=none] coordinates {
      ( 6.400e+01 , 1.805e-03 )
      ( 1.250e+02 , 2.143e-03 )
      ( 2.160e+02 , 1.828e-03 )
      ( 3.430e+02 , 2.107e-03 )
      ( 7.290e+02 , 2.827e-03 )
      ( 1.331e+03 , 3.737e-03 )
      ( 2.744e+03 , 8.121e-03 )
      ( 4.913e+03 , 1.296e-02 )
      ( 9.261e+03 , 2.619e-02 )
      ( 1.758e+04 , 5.602e-02 )
      ( 3.594e+04 , 1.297e-01 )
      ( 6.892e+04 , 2.598e-01 )
    };
    \addplot[dashed, color={green!60!black}, mark=none] coordinates {
      ( 6.400e+01 , 3.135e-02 )
      ( 1.250e+02 , 3.196e-02 )
      ( 2.160e+02 , 3.423e-02 )
      ( 3.430e+02 , 3.305e-02 )
      ( 7.290e+02 , 3.934e-02 )
      ( 1.331e+03 , 7.220e-02 )
      ( 2.744e+03 , 1.809e-01 )
      ( 4.913e+03 , 3.825e-01 )
      ( 9.261e+03 , 1.215e+00 )
      ( 1.758e+04 , 4.220e+00 )
      ( 3.594e+04 , 1.822e+01 )
      ( 6.892e+04 , 7.212e+01 )
    };
    \addplot[solid, color=green, mark=none] coordinates {
      ( 6.400e+01 , 1.507e-03 )
      ( 1.250e+02 , 2.024e-03 )
      ( 2.160e+02 , 1.956e-03 )
      ( 3.430e+02 , 2.358e-03 )
      ( 7.290e+02 , 2.870e-03 )
      ( 1.331e+03 , 3.672e-03 )
      ( 2.744e+03 , 7.497e-03 )
      ( 4.913e+03 , 1.292e-02 )
      ( 9.261e+03 , 2.494e-02 )
      ( 1.758e+04 , 5.315e-02 )
      ( 3.594e+04 , 1.299e-01 )
      ( 6.892e+04 , nan )
    };
    \addplot[dashed, color=green, mark=none] coordinates {
      ( 6.400e+01 , 4.044e-02 )
      ( 1.250e+02 , 4.236e-02 )
      ( 2.160e+02 , 4.189e-02 )
      ( 3.430e+02 , 4.218e-02 )
      ( 7.290e+02 , 4.275e-02 )
      ( 1.331e+03 , 7.038e-02 )
      ( 2.744e+03 , 1.581e-01 )
      ( 4.913e+03 , 3.617e-01 )
      ( 9.261e+03 , 6.802e-01 )
      ( 1.758e+04 , 1.342e+00 )
      ( 3.594e+04 , 2.952e+00 )
      ( 6.892e+04 , 6.389e+00 )
    };
    \addplot[solid, color={rgb,255:red,0;green,255;blue,255}, mark=none] coordinates {
      ( 6.400e+01 , 8.896e-04 )
      ( 1.250e+02 , 2.536e-03 )
      ( 2.160e+02 , 5.832e-03 )
      ( 3.430e+02 , 1.037e-02 )
      ( 7.290e+02 , 2.680e-02 )
      ( 1.331e+03 , 5.745e-02 )
      ( 2.744e+03 , 1.589e-01 )
      ( 4.913e+03 , 3.011e-01 )
      ( 9.261e+03 , 6.025e-01 )
      ( 1.758e+04 , 1.361e+00 )
      ( 3.594e+04 , 2.692e+00 )
      ( 6.892e+04 , 4.970e+00 )
    };
    \addplot[dashed, color={rgb,255:red,0;green,255;blue,255}, mark=none] coordinates {
      ( 6.400e+01 , 9.978e-04 )
      ( 1.250e+02 , 2.001e-03 )
      ( 2.160e+02 , 3.548e-03 )
      ( 3.430e+02 , 5.773e-03 )
      ( 7.290e+02 , 1.414e-02 )
      ( 1.331e+03 , 2.926e-02 )
      ( 2.744e+03 , 8.446e-02 )
      ( 4.913e+03 , 2.230e-01 )
      ( 9.261e+03 , 6.172e-01 )
      ( 1.758e+04 , 1.648e+00 )
      ( 3.594e+04 , 4.518e+00 )
      ( 6.892e+04 , 1.110e+01 )
    };
    \addplot[solid, color={rgb,255:red,255;green,0;blue,255}, mark=none] coordinates {
      ( 6.400e+01 , 7.779e-04 )
      ( 1.250e+02 , 2.476e-03 )
      ( 2.160e+02 , 5.525e-03 )
      ( 3.430e+02 , 1.024e-02 )
      ( 7.290e+02 , 2.907e-02 )
      ( 1.331e+03 , 5.676e-02 )
      ( 2.744e+03 , 1.572e-01 )
      ( 4.913e+03 , 3.027e-01 )
      ( 9.261e+03 , 6.055e-01 )
      ( 1.758e+04 , 1.345e+00 )
      ( 3.594e+04 , nan )
      ( 6.892e+04 , nan )
    };
    \addplot[dashed, color={rgb,255:red,255;green,0;blue,255}, mark=none] coordinates {
      ( 6.400e+01 , 2.695e-03 )
      ( 1.250e+02 , 5.225e-03 )
      ( 2.160e+02 , 8.877e-03 )
      ( 3.430e+02 , 1.166e-02 )
      ( 7.290e+02 , 3.186e-02 )
      ( 1.331e+03 , 4.134e-02 )
      ( 2.744e+03 , 1.261e-01 )
      ( 4.913e+03 , 2.388e-01 )
      ( 9.261e+03 , 6.864e-01 )
      ( 1.758e+04 , 1.715e+00 )
      ( 3.594e+04 , 4.738e+00 )
      ( 6.892e+04 , 1.166e+01 )
    };
    \addplot[dotted, color=black, mark=none] coordinates {
      ( 6.400e+01 , 1.395e-03 )
      ( 1.250e+02 , 1.510e-03 )
      ( 2.160e+02 , 1.807e-03 )
      ( 3.430e+02 , 2.076e-03 )
      ( 7.290e+02 , 2.841e-03 )
      ( 1.331e+03 , 3.676e-03 )
      ( 2.744e+03 , 7.549e-03 )
      ( 4.913e+03 , 1.359e-02 )
      ( 9.261e+03 , 2.501e-02 )
      ( 1.758e+04 , 5.362e-02 )
      ( 3.594e+04 , 1.339e-01 )
      ( 6.892e+04 , 3.121e-01 )
    };
    \addlegendentry[fill=white]{expl\_legacy}
    \addlegendentry[fill=white]{impl\_legacy}
    \addlegendentry[fill=white]{expl\_modern}
    \addlegendentry[fill=white]{impl\_modern}
    \addlegendentry[fill=white]{expl\_mkl}
    \addlegendentry[fill=white]{impl\_mkl}
    \addlegendentry[fill=white]{expl\_cholmod}
    \addlegendentry[fill=white]{impl\_cholmod}
    \addlegendentry[fill=white]{expl\_hybrid}
  \end{loglogaxis}
\end{tikzpicture}
        \caption{Heat transfer 3D, application}
        \label{fig:time_karolina_3d_apply}
    \end{subfigure}
    \caption{Execution time of preprocessing and application}
    \label{fig:time_karolina}
\end{figure*}

The comparison for heat transfer is shown in Fig.~\ref{fig:time_karolina}; for linear elasticity problems, the results are similar, thus we do not present them here.
In the top two graphs, 2D problems are measured, while in the bottom two, there are 3D problems. On the left we plotted the time of preprocessing, on the right you can see the time of application.
The explicit assembly parameters were set according to the recommendation in Table~\ref{tab:optimal_settings}.

In general, the fastest preprocessing of the dual operator provides PARDISO from the Intel MKL library using the implicit approach. Comparing it with CHOLMOD from the SuiteSparse library, MKL PARDISO is able to factorize subdomains from 2D and small 3D meshes about 2 times faster. For large 3D subdomains, differences are only marginal.

The preprocessing in the implicit GPU approach is only slightly slower than in implicit CHOLMOD.
The difference is caused only by extracting and copying the factors.
If MKL PARDISO allowed the extraction of factors, the implicit GPU preprocessing would copy the trend of the implicit MKL PARDISO preprocessing.

As the explicit preprocessing assembles the matrices $\tilde{\mF}_i$, it is naturally slower than implicit preprocessing.
A surprising exception is the explicit MKL PARDISO approach for 2D subdomains, which is faster than the implicit CHOLMOD approach.
This is caused by the very fast factorization for the 2D problems by MKL PARDISO and the efficient utilization of a sparsity pattern by the augmented incomplete factorization.

The assembly of the explicit $\tilde{\mF}_i$ by CUDA is very fast for moderately-sized 2D subdomains, where it takes only slightly longer than CHOLMOD preprocessing. In 3D, the time increases at most 4 times for up to 5000 DOFs.
In the case of small subdomains, the overhead of CUDA kernels is high.
For large subdomains, the slower assembly is caused by the increasing number of Lagrange multipliers in $\tilde{\mB}_i$.
The difference between legacy and modern CUDA is caused by the low performance of the sparse TRSM in modern CUDA.

The situation is different for subdomains from the 3D mesh. Factors are much denser, and the difference between factorization with MKL PARDISO and CHOLMOD is small.
Explicit assembly using CUDA is faster than the explicit approach with MKL PARDISO. Due to the higher density of factors and the use of dense TRSM, there is minimal difference between CUDA modern and CUDA legacy for subdomains below $12000 \approx 2^{13.5}$ DOFs.
With the increasing subdomain sizes (and increasing sparsity), the advantage of CUDA over the incomplete augmented factorization from MKL PARDISO is getting smaller.
Naturally, the ratio between these two approaches is highly dependent on the ratio between CPU and GPU performance and memory bandwidth.
The explicit approach with CHOLMOD is generally the slowest, since it does not utilize the sparsity pattern of $\tilde{\mB}_i$ and uses the CPU.

Moving on to the application, using the explicitly assembled $\tilde{\mF}_i$ on the GPU is generally the fastest (except for small subdomains where overheads dominate).
The improvement is caused by the dense matrix representation of $\mF$ and by the higher GPU memory bandwidth compared to the CPU.

The higher speed-up of explicit approaches for subdomains from the 2D mesh compared to speed-up for 3D mesh is caused by a better ratio between the number of DOFs and the number of Lagrange multipliers, i.e., $\tilde{\mF}_i$ is proportionally smaller compared to $\mK_i$ for the 2D mesh.

Contrary to the assembly, the application of $\tilde{\mF}_i$ is slightly faster with CUDA modern than with CUDA legacy. This is probably caused by better optimizations in the newer version.

There is no difference in the application between the two explicit CPU approaches. This is because they use the same matrices, and the library that assembles $\tilde{\mF}_i$ does not have an influence here.
Implicit application on CPU is again faster in MKL PARDISO than CHOLMOD for subdomains from the 2D mesh and small subdomains from the 3D mesh.

The hybrid approach (original GPU implementation) simply copies the trend of the MKL PARDISO assembly on the CPU and the explicit application on the GPU.

\subsection{Choosing the best dual operator approach}

\begin{figure*}
    \begin{subfigure}[b]{0.47\textwidth}
        \centering
        \input{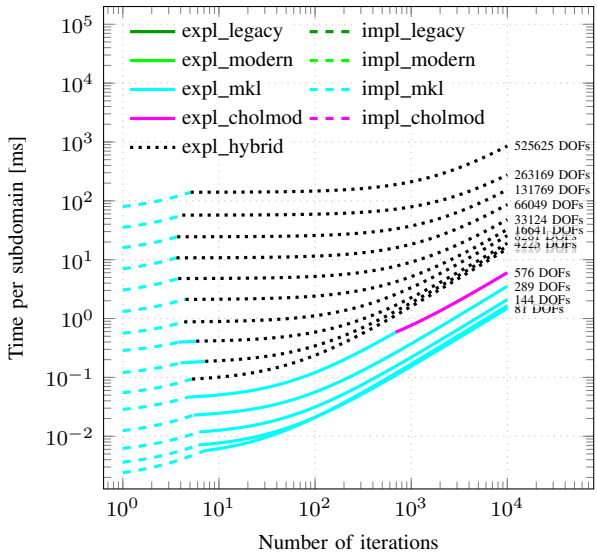}
        \caption{Heat transfer 2D}
        \label{fig:best_dualop_2d}
    \end{subfigure}
    \hfill
    \begin{subfigure}[b]{0.47\textwidth}
        \centering
        \input{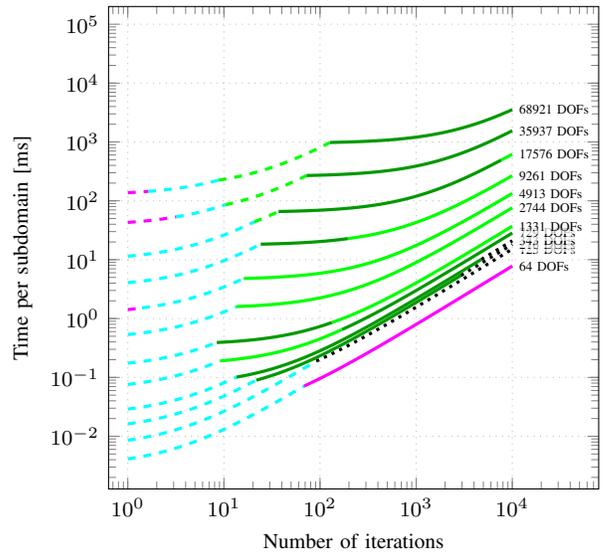}
        \caption{Heat transfer 3D}
        \label{fig:best_dualop_3d}
    \end{subfigure}
    \caption{Time-step time of the best dual operator}
    \label{fig:best_dualop}
\end{figure*}

In Fig.~\ref{fig:best_dualop}, we plot the overall time spent in the dual operator -- it shows the time of preprocessing and iterations summed together. For each subdomain size, the line denotes the time of only the best dual operator approach for the number of iterations given by the X-axis. The color and style of the line denote the approach used.

As can be seen, if the solver does only several iterations, the implicit approach with MKL PARDISO is the best option for both 2D and 3D subdomains.
As the number of iterations increases, the explicit approaches start to improve.
The amortization point (the number of iterations, where the explicit approach starts to be better than implicit) depends on problem dimensionality and subdomain size.

For 2D meshes, MKL PARDISO dominates.
For small subdomains, the CPU-only explicit approach is the best because of the high latency of CUDA.
For larger subdomains, the hybrid approach is optimal.
The MKL PARDISO domination is caused by the slow factorization in the CHOLMOD library, as discussed in the previous subsection.
If we had a library that provides factors with the same speed as MKL PARDISO, our accelerated approach for medium-size subdomains would be better according to the graphs in Fig.~\ref{fig:time_karolina_2d_update} (difference between the \textit{impl\_cholmod} and \textit{expl\_legacy} is smaller than the difference between \textit{impl\_mkl} and \textit{expl\_mkl}).

For 3D meshes, accelerated explicit approaches are better (CUDA legacy or CUDA modern).
About ten iterations are enough to make the acceleration beneficial for medium-sized domains (about 500 - 2000 DOFs).
With increasing the subdomain size, the number of iterations until amortization increases due to the longer preprocessing time.

The graphs in Fig.~\ref{fig:best_dualop} can be used to decide whether acceleration is beneficial and which dual operator approach should be used for best timing results.

\begin{figure*}
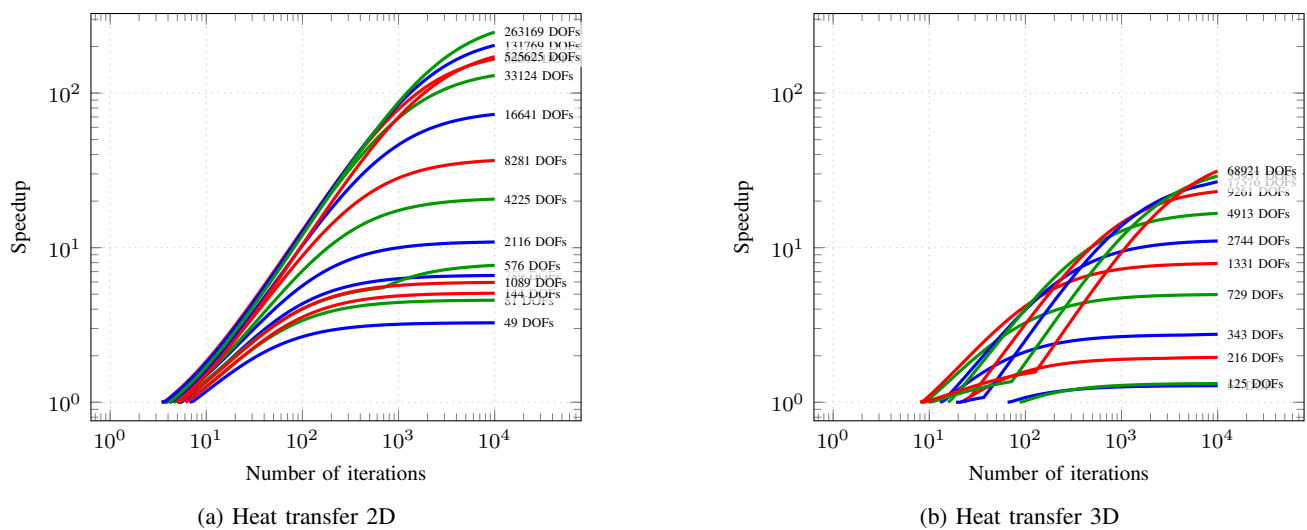

    \begin{subfigure}[b]{0.47\textwidth}
        \centering
        \input{tikz/speedup/karolina-heat_transfer_2d-TRIANGLE3.tex}
        \caption{Heat transfer 2D}
        \label{fig:speedup_2d}
    \end{subfigure}
    \hfill
    \begin{subfigure}[b]{0.47\textwidth}
        \centering
        \input{tikz/speedup/karolina-heat_transfer_3d-TETRA4.tex}
        \caption{Heat transfer 3D}
        \label{fig:speedup_3d}
    \end{subfigure}
    \caption{Speedup of the best dual operator relative to the implicit CPU approach}
    \label{fig:speedup}
\end{figure*}

Fig.~\ref{fig:speedup} shows the maximal speedup for a given number of iterations and subdomain sizes compared to the implicit CPU approach with Intel MKL.
The beginning of each line denotes the number of iterations in which another approach begins to be beneficial. The lines (speedup) go up with the increasing number of iterations until they reach the limit determined by the speedup of a single application shown in Fig.~\ref{fig:time_karolina_2d_apply} and~\ref{fig:time_karolina_3d_apply}.
The visible changes in the trends are caused by a change in the optimal dual operator approach (e.g., from implicit to explicit).

Fig.~\ref{fig:speedup} also shows that using the correct approach can have an enormous impact on the computation time.
For 2D meshes with subdomains with more than 10,000 DOFs and 100 PCPG iterations, the FETI solver can be more than 10 times faster if the explicit approach is used.
For ill-conditioned problems with hundreds of iterations, the explicit approach reaches an overall speed-up of about 100.
In general, the best speed-up was observed for subdomains with 100,000--300,000 DOFs.
For 3D meshes, the speed-up is substantially smaller due to the mentioned ratio between the number of Lagrange multipliers and the size of subdomains, which is better for 2D meshes.
A speed-up of 4 was achieved for subdomains with around 2,000 DOFs and 100 PCPG iterations.
For ill-conditioned problems with hundreds of iterations, a speed-up of about 10 was observed for subdomains with 4,000--20,000 DOFs.

The graphs in Fig.~\ref{fig:speedup} can be used to estimate the expected speedup of the FETI solver for given subdomain sizes and the number of PCPG iterations.

\section{Conclusion}

The paper presents an approach for the acceleration of explicit assembly of the FETI dual operator $\mF = \mB \mK^+ \mB^\top$ using Nvidia A100.
Despite the operator being composed of sparse matrices, we have shown that modern Nvidia accelerators can be successfully used.

Large differences were observed between major versions of CUDA libraries and between various parameters of the assembly process.
Based on exhaustive measurements of all possible configurations on matrices generated by heat transfer and linear elasticity physics, the paper suggests optimal settings.
With the recommended settings, one can expect the best time-to-solution among all the available options.

Using Karolina GPU node with two 64-core AMD Zen3 CPUs and 8 Nvidia A100 GPUs, for problems that need hundreds of iterations of the FETI solver, we reach a speedup of up to 100 for 2D meshes with the hybrid explicit approach, and for 3D meshes the speedup is up to 10 with the explicit GPU approach.
The amortization point was between 5 and 10 iterations for 2D meshes and between 10 and 100 iterations for 3D meshes.

In future work, our aim is to test our approach on AMD and Intel GPU accelerators, and also on the newest Nvidia Grace-Hopper.
Additionally, we would like to develop an algorithm that is able to better utilize the sparsity pattern of matrices during the explicit assembly process on GPUs.

\section*{Acknowledgement}
This work was supported by the EUPEX project. This project has received funding from the European High-Performance Computing Joint Undertaking (JU) under grant agreement No 101033975. The JU receives support from the European Union’s Horizon 2020 research and innovation programme and France, Germany, Italy, Greece, United Kingdom, Czech Republic, Croatia. And by the Ministry of Education, Youth and Sports of the Czech Republic through the e-INFRA CZ (ID:90254).

\bibliographystyle{IEEEtran}
\bibliography{bib}

\end{document}